\documentclass[11pt,twoside]{article}
\usepackage{asp2010}

\resetcounters

\def \NII {[N{\sc ii}]6584~\AA}
\def \ha {H$\alpha$}

\def \vhel{\ifmmode{~V_{{\rm HEL}}}\else{~$V_{{\rm HEL}}$}\fi}
\def\kms{\ifmmode{~{\rm km\,s}^{-1}}\else{~km s$^{-1}$}\fi}

\begin{document}

\title{Abell 41: nebular shaping by a binary central star?}
\author{D. Jones$^1$, M. Lloyd$^1$, M. Santander-Garc\'ia$^{2,3,4}$, J.A. L\'opez$^5$, J. Meaburn$^1$, D.L. Mitchell$^1$, T.J. O'Brien$^1$, D. Pollacco$^6$, M.M. Rubio-D\'iez$^{7,2}$ and N.M.H. Vaytet$^8$
\affil{$^1$Jodrell Bank Centre for Astrophysics, University of Manchester, UK}
\affil{$^2$Isaac Newton Group of Telescopes, Santa Cruz de La Palma, Spain}
\affil{$^3$Instituto de Astrof\'isica de Canarias, Tenerife, Spain}
\affil{$^4$Departamento de Astrof\'isica, Universidad de La Laguna, Spain}
\affil{$^5$Instituto de Astronom\'ia, Universidad Nacional Aut\'onoma de M\'exico, M\'exico}
\affil{$^6$Astrophysics Research Centre, Queen's University Belfast, UK}
\affil{$^7$Departamento de Astrof\'isica, Centro de Astrobiolog\'ia, CSIC-INTA, Spain}
\affil{$^8$Service d'Astrophysique, CEA/DSM/IRFU/SAp, France}}

\begin{abstract}
We present the first detailed spatio-kinematical analysis and modelling of the planetary nebula Abell~41, which is known to contain the well-studied close-binary system MT Ser.  This object represents an important test case in the study of the evolution of planetary nebulae with binary central stars as current evolutionary theories predict that the binary plane should be aligned perpendicular to the symmetry axis of the nebula.

Longslit observations of the \NII\ emission from Abell~41 were obtained using the Manchester Echelle Spectrometer on the 2.1-m San Pedro M\'artir Telescope.   These spectra, combined with deep, narrowband imagery acquired using ACAM on the William Herschel Telescope, were used to develop a spatio-kinematical model of \NII\ emission from Abell~41.  The best fitting model reveals Abell~41 to have a waisted, bipolar structure with an expansion velocity of $\sim$40\kms{} at the waist. The symmetry axis of the model nebula is within 5$^\circ$ of perpendicular to the orbital plane of the central binary system.  This provides strong evidence that the close-binary system, MT Ser, has directly affected the shaping of its host nebula, Abell~41.

\end{abstract}

\section{Introduction}
\label{sec:intro}

Abell~41 (PN~G009.6+10.5, \mbox{$\alpha = 17^h29^m02.03^s$}, \mbox{$\delta = -15^\circ13'04.4''$} J2000), discovered by \cite{abell66}, was classified by \citet{bond90} as elliptical under the classification scheme of \citet{balick87}.  However, deeper \ha{}$+$\NII\ imagery reveals ``that the nebular morphology exhibits an `H' shape with the addition of fainter material forming a continuous loop'' \citep{pollacco97}.

Photometric analysis of the central star of the planetary nebula (CSPN), MT Ser, revealed it to be a close binary, showing minima at regular intervals of $2^h43^m$ \citep{grauer83}.  \citet{bruch01} confirmed the binary nature of MT Ser but were unable to accurately determine its orbital parameters because they found two different models which fit the observed data. (a) The binary consists of a hot sub-dwarf and a less evolved secondary, in which case the period is $2^h43^m$ and the variations are due to a reflection effect (inclination, $i = 42.52^\circ \pm 1.73 ^\circ$).  (b) The binary consists of two evolved, hot sub-dwarfs with a period of $5^h26^m$ where the variability results from partial eclipses and ellipsoidal variations ($i = 65.7^\circ \pm 0.9^\circ$).  They determined the optimum parameters for each model, but concluded that only radial velocity observations would be able to distinguish between the two.  Subsequent observation and modelling by \citet{shimanskii08} confirmed the presence of two sub-dwarf components, but gave no independent confirmation of the orbital inclination.  Only the second model of \citeauthor{bruch01} (\citeyear{bruch01}, $i = 65.7^\circ \pm 0.9^\circ$) is consistent with photometric observations and the detection of two hot sub-dwarf central stars, indicating that this is the most reliable model of the CSPN.

In these proceedings we present a spatio-kinematic model, derived from longslit spectroscopy and narrowband imaging, with the aim of understanding the relationship between the nebula and MT Ser.  A more complete discussion of this work can be found in \citep{jones10b}.

\begin{figure}
\centering
\includegraphics{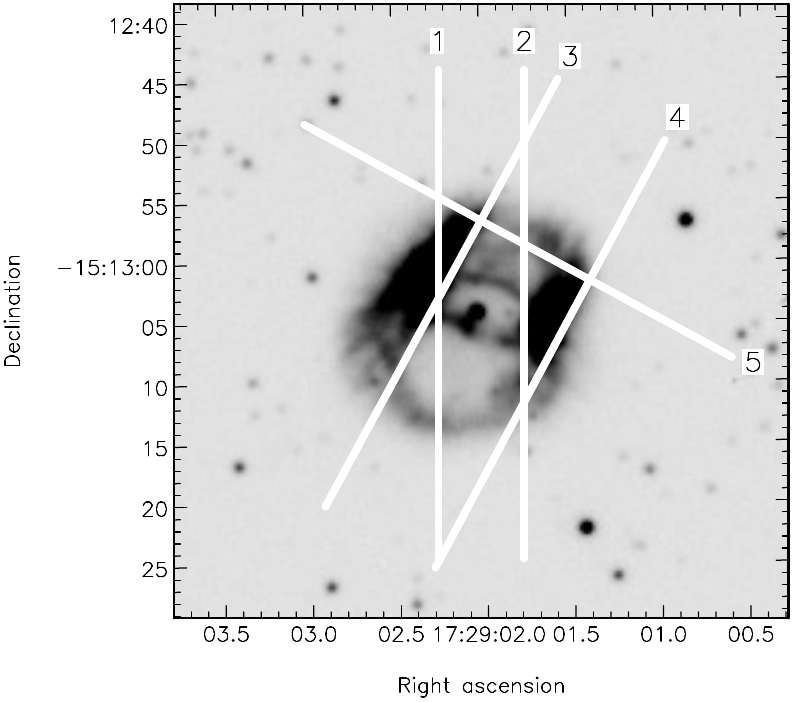}
\caption{A deep ACAM-WHT image of A~41 in the light of \NII\ showing the positions of the 5 MES-SPM longslits.}
\label{fig:slitimage}
\end{figure}

\section{Spatio-kinematic modelling}
\label{sec:modelling}

Spatially-resolved, high spectral-resolution echelle spectroscopy of the \NII\ line acquired using the Manchester Echelle Spectrograph on the 2.1-m San Pedro M\'artir telescope (MES-SPM), in combination with deep narrowband imagery obtained using ACAM on the 4.2-m William Herschel Telescope (WHT-ACAM), have been used to develop a spatio-kinematic model of A~41.  The modelling was performed using \textsc{shape} \citep{steffen06}.  Further details of the observations and modelling process are found in \citet{jones10b}.

\begin{figure}
\centering
\includegraphics{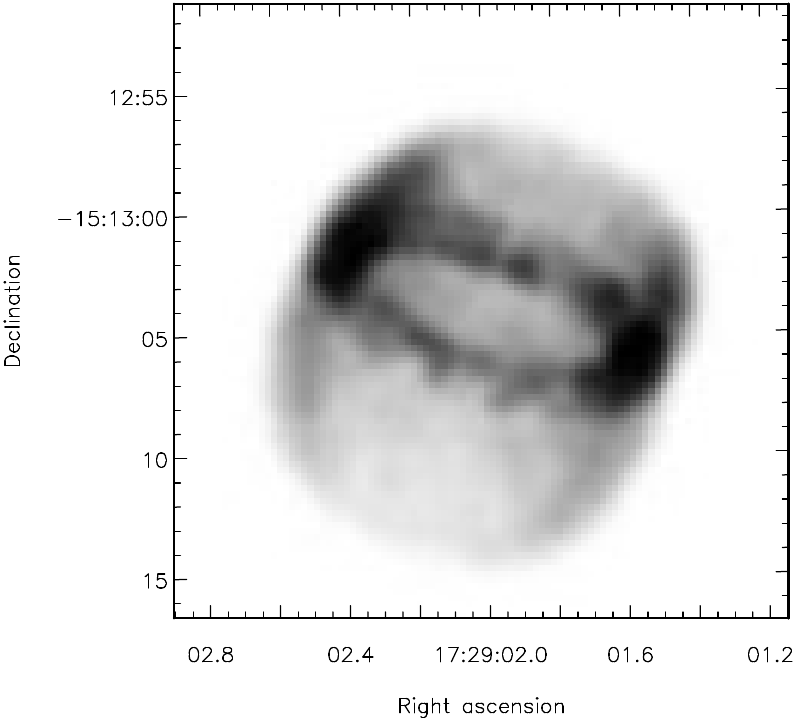}
\caption{The synthetic \textsc{shape} model for the \NII\ emission from A~41.}
\label{fig:modelimage}
\end{figure}

This best fit model comprises a bipolar shell waisted by an equatorial ring with an expansion velocity of $\sim$40\kms{}.  The model nebula is slightly asymmetric in that the northern lobe is shortened by 15\%, has a narrower opening angle and has a slight shear with respect to its southern counterpart. No symmetric model could be found which reproduced the observed PV arrays. The nebular inclination angle, as defined by the un-sheared southern lobe, is determined to be $66^\circ \pm 5^\circ$, in excellent agreement with the inclination of the CSPN (see section \ref{sec:intro}).  The model nebula is shown at the observed orientation in Figure \ref{fig:modelimage}.  The synthetic position-velocity arrays can be found, along with their observed counterparts, in \citet{jones10b}.

\subsection{Systemic velocity and kinematical age}

Comparison of synthetic model spectra to their observed counterparts provides an unambiguous measure of the nebular systemic heliocentric radial velocity ($V_{sys}$), unaffected, for example, by brightness variations or nebular asymmetry \citep{jones10a}.  Using our best-fit model, $V_{sys}$ is determined to be 30 $\pm$ 5 \kms\ in good agreement with the value of 30 \kms\ determined by \citet{beaulieu99}.  

The nebular expansion velocity, determined by the kinematical modelling, can be used to calculate a kinematical age for the nebula.  This, however, requires the distance to the nebula to be known.  The distance to A~41 is a matter of some debate with values in the literature ranging from $\sim$1 kpc \citep{grauer83} up to 9.0 $\pm$ 0.4 kpc \citep{shimanskii08}, this probably results from the notorious variation in results from different methods of distance determination (see e.g., \citealp{gurzadyan97}).  Therefore, rather than favour one particular distance estimate over another we quote a kinematical age per kiloparsec of $\sim800$ years kpc$^{-1}$.

\section{Conclusions}

Using high-resolution longslit spectroscopy and deep imaging, a spatio-kinematical model of A~41 has been developed which clearly shows that the nebula is aligned with the binary central system exactly as predicted by current theories of PN shaping by binary CSPN.  This is only the second nebula to have this link observationally constrained (after A~63, \citealp{mitchell07b}).  The kinematical data confirm A~41 exhibits a waisted, bipolar structure, with some small deviations from perfect axisymmetry.  The presence of an equatorial ring is also confirmed, adding further weight to the link between ring structures and central binary stars as commented on by \citeauthor{miszalski09b} (\citeyear{miszalski09b}, \citeyear{miszalski10}) and \citet{lopez10}.

Further kinematical investigations, such as this and others presented in these proceedings (e.g. \citealp{huckvale10}, \citealp{tyndall10}), coupled with in-depth studies of the CSPN of other PNe with confirmed close-binary CSPN, are necessary to investigate the full extent of the influence of central star binarity on PN nebular shaping.  Only once a significant statistical sample has been acquired can generalisations be made about the role of CSPN binarity in PN evolution. 

\acknowledgements 
Based on observations made with the WHT operated on the island of La Palma by the ING at the Observatorio del Roque de los Muchachos of the IAC.  We would like to thank the staff of the San Pedro M\'artir Observatory.

\bibliography{literature.bib}

\begin{thebibliography}{}
\expandafter\ifx\csname natexlab\endcsname\relax\def\natexlab#1{#1}\fi
\expandafter\ifx\csname url\endcsname\relax
  \def\url#1{\texttt{#1}}\fi
\expandafter\ifx\csname urlprefix\endcsname\relax\def\urlprefix{URL }\fi
\providecommand{\eprint}[2][]{\url{#2}}

\bibitem[{Abell \& Goldreich(1966)}]{abell66}
Abell, G.~O., \& Goldreich, P. 1966, PASP, 78, 232

\bibitem[{Balick(1987)}]{balick87}
Balick, B. 1987, AJ, 94, 671

\bibitem[{{Beaulieu} et~al.(1999){Beaulieu}, {Dopita}, \&
  {Freeman}}]{beaulieu99}
{Beaulieu}, S.~F., {Dopita}, M.~A., \& {Freeman}, K.~C. 1999, ApJ, 515, 610

\bibitem[{Bond \& Livio(1990)}]{bond90}
Bond, H.~E., \& Livio, M. 1990, ApJ, 335, 568

\bibitem[{Bruch et~al.(2001)Bruch, Vaz, \& Diaz}]{bruch01}
Bruch, A., Vaz, L.~P.~R., \& Diaz, M.~P. 2001, A\&A, 377, 898

\bibitem[{Grauer \& Bond(1983)}]{grauer83}
Grauer, A.~D., \& Bond, H.~E. 1983, ApJ, 271, 259

\bibitem[\protect\citeauthoryear{{Gurzadyan}}{{Gurzadyan}}{1997}]{gurzadyan97}
{Gurzadyan} G.~A.,  1997, {The Physics and Dynamics of Planetary Nebulae}.
Springer

\bibitem[{{Huckvale} et~al.(2010){Huckvale}, {Prouse}, {Jones}, {Lloyd} \& {Pollacco}}]{huckvale10}
{Huckvale}, L., {Prouse}, B., {Jones}, D., {Lloyd}, M. \& {Pollacco}, D.L. 2010, in Asymmetric Planetary Nebulae 5 edited by {Zijlstra}, A.A., {McDonald}, I. \& {Lagadec}, E.

\bibitem[{{Jones} et~al.(2010{\natexlab{a}}){Jones}, {Lloyd}, {Mitchell},
  {Pollacco}, {O'Brien}, \& {Vaytet}}]{jones10a}
{Jones}, D., {Lloyd}, M., {Mitchell}, D.~L., {Pollacco}, D.~L., {O'Brien},
  T.~J., \& {Vaytet}, N.~M.~H. 2010{\natexlab{a}}, MNRAS, 401, 405.

\bibitem[{{Jones} et~al.(2010{\natexlab{b}}){Jones}, {Lloyd},
  {Santander-Garc{\'{\i}}a}, {L{\'o}pez}, {Meaburn}, {Mitchell}, {O'Brien},
  {Pollacco}, {Rubio-D{\'{\i}}ez}, \& {Vaytet}}]{jones10b}
{Jones}, D., {Lloyd}, M., {Santander-Garc{\'{\i}}a}, M., {L{\'o}pez}, J.~A.,
  {Meaburn}, J., {Mitchell}, D.~L., {O'Brien}, T.~J., {Pollacco}, D.,
  {Rubio-D{\'{\i}}ez}, M.~M., \& {Vaytet}, N.~M.~H. 2010{\natexlab{b}}, MNRAS, accepted for publication, ArXiV eprint \eprint{1006.5873}
  
  \bibitem[{{L\'opez} et~al.(2010){L\'opez}, {Garc\'ia-D\'iaz}, {Richer}, {Lloyd} \& {Meaburn}}]{lopez10}
{L\'opez}, J.A., {Garc\'ia-D\'iaz}, M.T., {Richer}, M., {Lloyd}, M. \& {Meaburn}, J. 2010, in Asymmetric Planetary Nebulae 5 edited by {Zijlstra}, A.A., {McDonald}, I. \& {Lagadec}, E.

\bibitem[{{Miszalski} et~al.(2009){Miszalski}, {Acker}, {Parker}, \&
  {Moffat}}]{miszalski09b}
{Miszalski}, B., {Acker}, A., {Parker}, Q.~A., \& {Moffat}, A.~F.~J. 2009,
  A\&A, 505, 249.
  
  \bibitem[{{Miszalski} et~al.(2010){Miszalski}, {Corradi}, {Jones}, {Santander-Garc\'ia}, {Rodr\'iguez-Gil} \& {Rubio-D\'iez}}]{miszalski10}
{Miszalski}, B., {Corradi}, R.L.M., {Jones}, D., {Santander-Garc\'ia}, M., {Rodr\'iguez-Gil}, P., \& {Rubio-D\'iez}, M.M. 2010, in Asymmetric Planetary Nebulae 5 edited by {Zijlstra}, A.A., {McDonald}, I. \& {Lagadec}, E.

\bibitem[{Mitchell et~al.(2007)Mitchell, Pollacco, O'Brien, Bryce, Lopez,
  Meaburn, \& Vaytet}]{mitchell07b}
Mitchell, D.~L., Pollacco, D., O'Brien, T.~J., Bryce, M., Lopez, J.~A.,
  Meaburn, J., \& Vaytet, N.~M.~H. 2007, MNRAS, 374, 1404

\bibitem[{Pollacco \& Bell(1997)}]{pollacco97}
Pollacco, D., \& Bell, S.~A. 1997, MNRAS, 284, 32

\bibitem[{{Shimanskii} et~al.(2008){Shimanskii}, {Borisov}, {Sakhibullin}, \&
  {Sheveleva}}]{shimanskii08}
{Shimanskii}, V.~V., {Borisov}, N.~V., {Sakhibullin}, N.~A., \& {Sheveleva},
  D.~V. 2008, Astronomy Reports, 52, 479

\bibitem[{Steffen \& Lopez(2006)}]{steffen06}
Steffen, W., \& Lopez, J.~A. 2006, RMxAA, 42, 99

\bibitem[{{Tyndall} et~al.(2010){Tyndall}, {Jones}, {Lloyd}, {O'Brien}, {Pollacco} \& {Mitchell}}]{tyndall10}
{Tyndall}, A., {Jones}, D., {Lloyd}, M., {O'Brien}, T.J., {Pollacco}, D.L. \& {Mitchell}, D.L. 2010, in Asymmetric Planetary Nebulae 5 edited by {Zijlstra}, A.A., {McDonald}, I. \& {Lagadec}, E.


\end{thebibliography}

\end{document}